\documentclass[12pt]{iopart}

\usepackage{graphicx}
\usepackage{color}

\begin{document}

\title[Fused silica vs. crystalline quartz]{On the mechanical quality factors of cryogenic test masses from fused silica and crystalline quartz}

\author{A Schroeter$^1$, R Nawrodt$^1$, R Schnabel$^3$, S Reid$^4$, I Martin$^4$, S Rowan$^4$, C Schwarz$^1$, T Koettig$^1$, R Neubert$^1$, M Th\"urk$^1$, W Vodel$^1$, A T\"unnermann$^2$, K Danzmann$^3$, P Seidel$^1$}

\address{$^1$ Institut f\"ur Festk\"orperphysik, Friedrich-Schiller-Universit\"at Jena, Helmholtzweg~5, D-07743~Jena, Germany}

\address{$^2$ Institut f\"ur Angewandte Physik, Friedrich-Schiller-Universit\"at Jena, Max-Wien-Platz~1, D-07743~Jena, Germany}

\address{$^3$ Max-Planck-Institut f\"ur Gravitationsphysik (Albert-Einstein-Institut) and Leibniz Universit\"at Hannover, Callinstr.~38, D-30167~Hannover, Germany}

\address{$^4$ SUPA, University of Glasgow, Glasgow, G128QQ, UK}

\ead{anja.zimmer@uni-jena.de}

\begin{abstract}
Current interferometric gravitational wave detectors (IGWDs) are
operated at room temperature with test masses made from fused
silica. Fused silica shows very low absorption at the laser
wavelength of 1064\,nm. It is also well suited to realize low
thermal noise floors in the detector signal band since it offers
low mechanical loss, i.\,e. high quality factors (Q factors) at
room temperature. However, for a further reduction of thermal
noise, cooling the test masses to cryogenic temperatures
may prove an interesting technique. Here we compare the results of
Q factor measurements at cryogenic temperatures of acoustic
eigenmodes of test masses from fused silica and its crystalline
counterpart. Our results show that the mechanical loss of fused
silica increases with lower temperature and reaches a maximum at
30\,K for frequencies of slightly above 10\,kHz. The losses of
crystalline quartz generally show lower values and even fall below
the room temperature values of fused silica below 10\,K. Our
results show that in comparison to fused silica, crystalline
quartz has a considerably narrower and lower dissipation peak on
cooling and thus has more promise as a test mass material for
IGDWs operated at cryogenic temperatures. The origin of the
different Q factor versus temperature behavior of the two
materials is discussed.
\end{abstract}


\maketitle                   




\renewcommand{\leftmark}
{Schroeter et al.: Quality factors of fused silica and crystalline quartz}

\section{Introduction}
\label{introduction}

The current interferometric gravitational wave detectors (IGWDs)
\cite{Sau1994,Hou2005,Auf2005} designed to search for
gravitational radiation produced by astrophysical sources are
among the most sensitive measurement devices ever built. These
instruments operate by sensing changes in the relative positions
of test-mass mirrors suspended as pendulums. Noise spectral
densities for differential strain measurements lower than
10$^{-22}\mbox{m}/\sqrt{\mbox{Hz}}$ at frequencies between 70\,Hz
and about 400\,Hz have been achieved \cite{LIGO2006}. Future
generations of gravitational wave detectors (GWDs) aim for even
lower noise spectral densities. To achieve this goal noise sources
that produce displacements of the interferometer's test mass
surfaces in the direction of the optical axis must be further
reduced because their influence mimics that of gravitational
waves. Noise sources that exist due to a non-zero temperature are
called \emph{thermal} noise sources. One source for thermal noise
originates from thermally driven mechanical vibrations of the test
masses themselves.\\
Until relatively recently the thermal noise from test mass and
suspensions was modelled by treating the resonant modes of the
systems as damped harmonic oscillators. Applying the
fluctuation-dissipation theorem \cite{Cal1951,Cal1952}, the power
spectral density of thermal displacement noise for each mode may
be found, and the total thermal noise calculated by considering
the laser beam reflected from the front face of a test mass as
sensing the incoherent sum of the thermal displacements in the
'tails' of the test mass resonant modes (see for example
\cite{Gil1995}).\\
However a method for calculating thermal noise was developed by
Levin \cite{Lev1998} which showed that the modal approach is
accurate only when the off-resonance thermal noise from each mode
is uncorrelated. In a system having spatially inhomogeneous
mechanical dissipation, correlations exist between the noise from
the modes, and the resulting thermal noise can be greater or
smaller than a modal treatment would suggest
\cite{Lev1998,Yam2002}. Using Levin's approach the power spectral
density $ S_{x}\left(f\right)$ of thermally excited displacement
of the front face of a test mass mirror may be written as

\begin{equation}
    S_{x}\left(f\right) = \frac{2 k_{B} T}{\pi^{2} f^{2}} \frac{W_{diss}}{F_{0}^{2}}, \label{Sx2}
\end{equation}

where $f$ is frequency, T is the temperature, $k_{B}$ is the
Boltzmann constant and $W_{diss}$ is the power dissipated when a
notional oscillatory force of peak magnitude $F_0$ acting on the
face of a test mass mirror has an associated pressure of spatial
profile identical to that of the laser beam used to sense the
mirror displacement. No general formula can be given if the loss
is inhomogeneous, however if the loss is homogeneous, and in the
case where a test mass can be approximated as semi-infinite in
dimension (compared with the diameter of a sensing laser beam)
then it can be shown that the power spectral density of thermal
noise of a test mass substrate having mechanical loss
$\phi_{substrate}(f) $ is given by

\begin{equation}
S_{x}\left(f\right) = \frac{2 k_{B} T}{\pi
f}\frac{1-\sigma^{2}}{\sqrt{2\pi} E r_{0} } \phi_{substrate}(f)
\end{equation}
where $E$  and $\sigma$ are the Young's modulus and Poisson's
ratio of the material and $r_{0}$ is the radius of the laser beam
at a point where the intensity has fallen to 1/e of its maximum
\cite{Liu2002}.

A direct measurement of the relevant mechanical losses at the
frequencies of the detection band of IGWDs is often impracticable
as the losses are typically extremely low in this region. A more
feasible approach is to determine the mechanical losses at the
resonances of the test masses. At a resonant frequency, the
reciprocal of the mechanical loss factor is given by the quality
factor, or Q factor, of the oscillation.\\
According to the fluctuation-dissipation theorem, the thermal
noise can be reduced by decreasing the temperature of the test
masses, thus cooling the test masses in IGWDs is of interest as a
technique for reducing thermal noise. See for example
\cite{Uch2004}. The mechanical loss of the test masses generally
depends on temperature as well as on frequency, thus for a given
material the temperature dependence of mechanical loss is an
important parameter to study.

For single relaxation processes with merely one relaxation time
$\tau$, the tangent of the mechanical loss $\phi$ can be expressed
in the following way \cite{Zen1948,Now1972},
\begin{equation}
    \tan \left( \phi \right) \: = \: \Delta \: \frac{2\pi \cdot f \cdot \tau}{1+ \left(2\pi \cdot f \cdot \tau \right)^{2}}\,,
    \label{zener}
\end{equation}
where $\Delta$ is the so-called relaxation strength.  For most
processes the relaxation time $\tau$ strongly varies with
temperature and also the relaxation strength has a temperature
dependent component for some processes \cite{Now1972}.

Current interferometric gravitational wave detectors are operated
at room temperature with test masses made from fused silica. Fused
silica shows very low absorption at the laser wavelength of
1064\,nm \cite{Hil2006}. It is also well suited to realize low
thermal noise floors in the detector signal band since it offers
low mechanical loss, i.\,e. high quality factors (Q factors)  of
up to $2\times10^{8}$ at room temperature \cite{Age2004}. However,
for a further reduction of thermal noise cooling of the test
masses to cryogenic temperatures may be of interest.
Unfortunately, it has been found that Q factors of fused silica
samples significantly decrease with decreasing temperatures. Early
measurements used the so-called `resonant piezoelectric method'
and found a damping peak of fused silica at about 35\,K
\cite{Fin1954} measured at a resonant frequency of 50\,kHz. Other
early works were performed at rather high frequencies far above
the gravitational wave detection band ($>$\,500\,MHz)
\cite{Hun1974}. More recent investigations used Brillouin
spectroscopy and concentrated on small samples ($<$\,1\,cm$^3$). Q
factors of the order 10$^3$ were achieved \cite{Tie1992,Vac2005}.
\emph{Crystalline} quartz also offers very low absorption at the IGWD wavelength of 1064\,nm \cite{Bra2001}.
Q factor measurements on crystalline quartz samples were performed in \cite{Bom1956,Kin1958,Fra1964}. A much
richer variation of Q factor values with temperature was found. The highest Q factor value observed in those
works was slightly below 10$^7$ measured on a 2$\times$2$\times$2.5\,cm$^3$ natural quartz crystal at a
5\,MHz resonance \cite{Fra1964}.

In this work we reanalyze the behavior of fused silica and
crystalline quartz Q factors using the method of ``cryogenic
resonant acoustic spectroscopy of bulk materials'' \cite{Zim2007}
using methods based on those described in \cite{Bra1986,MLGHDG78}.
In this method excitation and readout is realized without
mechanical contact providing high measurement accuracies. We used
artificially produced samples with a high grade of purity allowing
for the resolution of narrow damping structures in the temperature
spectrum. Our data shows a greatly increased temperature
resolution compared with previous measurements. We also used
rather large samples having an increased relevance for test mass
applications in gravitational wave detectors and, for crystalline
quartz, observed Q factors of up to 7$\times$10$^7$ at a 11.5\,kHz
resonance.

\section{Measurement of Q factors}
\label{experimental}

The measurements of the Q factors were performed in a custom made cryostat. Details concerning the
measurement system can be found in the work of Nawrodt et al. \cite{Naw2006,Naw2007} and Zimmer et al.
\cite{Zim2007}. In the following, the measuring process is roughly outlined. The mechanical loss at a
resonant frequency was determined by cryogenic resonant acoustic (CRA) spectroscopy which is an amplitude
ring-down method. For this purpose the substrate was excited to resonant vibrations by means of a comb
capacitor which avoids additional losses due to contact to the substrate. After removing the  driving
electric field the subsequent ring-down of the vibrational amplitude was recorded by a Michelson-like
interferometer. From the ring-down time the Q factor was determined. To minimize additional losses due to
clamping, the substrate was kept in a loop of wire made of tungsten resulting in a pendulum. For
reduction of gas damping the probe chamber was evacuated to a vacuum of better than $10^{-3}$ Pa.\\
With this set-up, measurements at temperatures between 5\,K and
325\,K can be realized. For cylindrical samples of isotropic
materials the Q factor measured is insensitive to the rotation of
the substrate in the suspension. It is plausible that the mode
shape is self-aligned with respect to the contact points of the
suspension. In anisotropic materials the mode shape is aligned to
the coordinate system of the inner structure of the material
\cite{Log1991}. Thus, by rotating the substrate in the wire loop
the Q factor varies \cite{Hei2007}. The position with the highest Q factor at
room temperature was adopted for all of the measurement cycles in
our investigations.

\section{Results and Discussion}
\label{results}

\begin{figure}[htb]
\centering
\includegraphics[width=.8\textwidth]{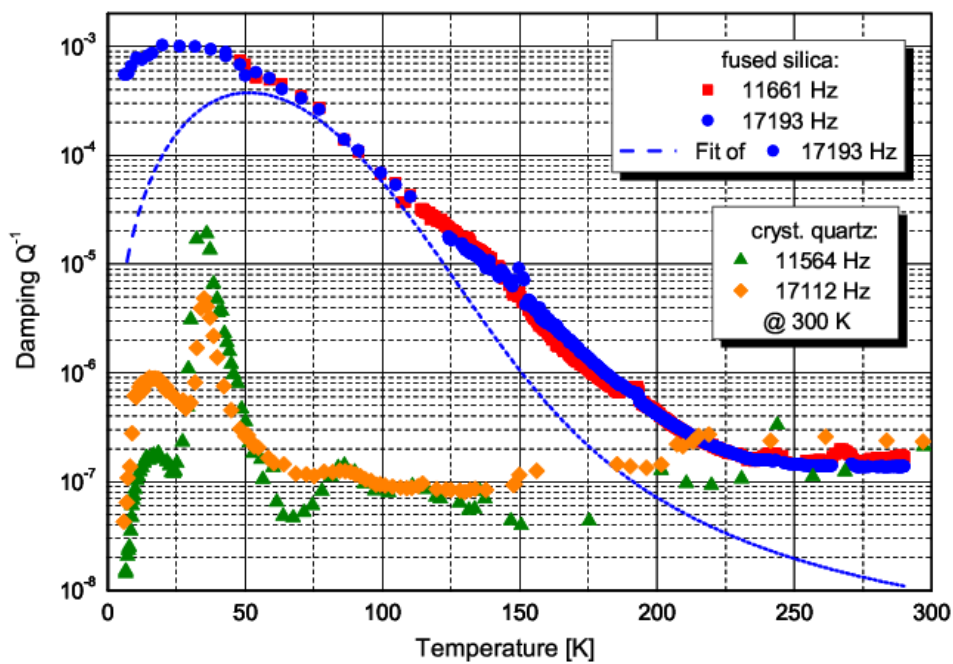}
\caption{Reciprocal of the measured Q factors versus temperature for a fused silica and crystalline quartz
substrate both 76.2\,mm in  diameter and 12\,mm thick. The dashed line represents a theoretical model for a
dominant loss mechanism at intermediate temperatures around 80\,K which was fitted to the measured data of
fused silica at 17193 Hz. Details of the fit parameters are given in the text. } \label{fig:fusedsilica}
\end{figure}

\begin{figure}[htb]
a)~\includegraphics[width=.2\textwidth]{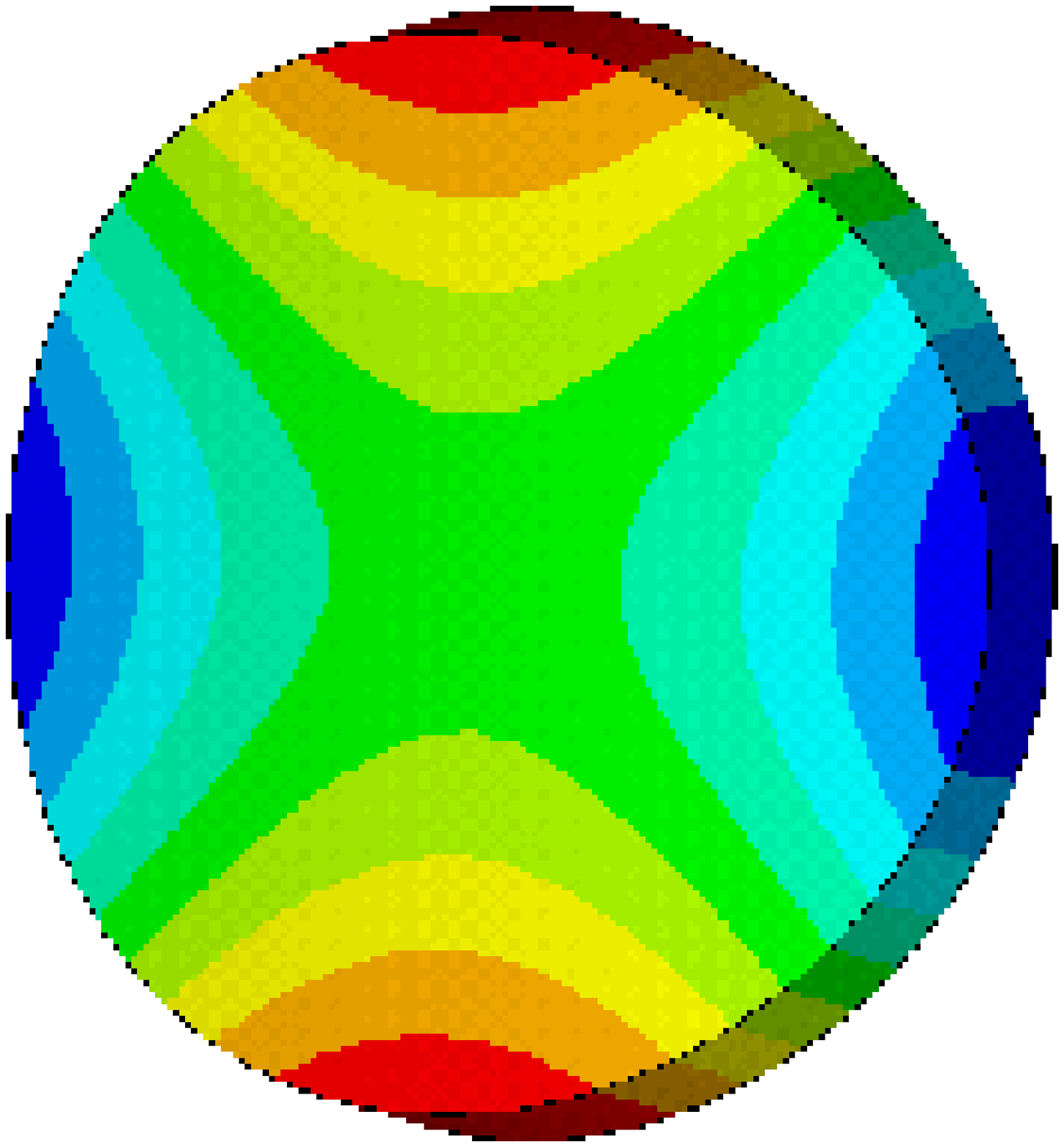} b)~\includegraphics[width=.2\textwidth]{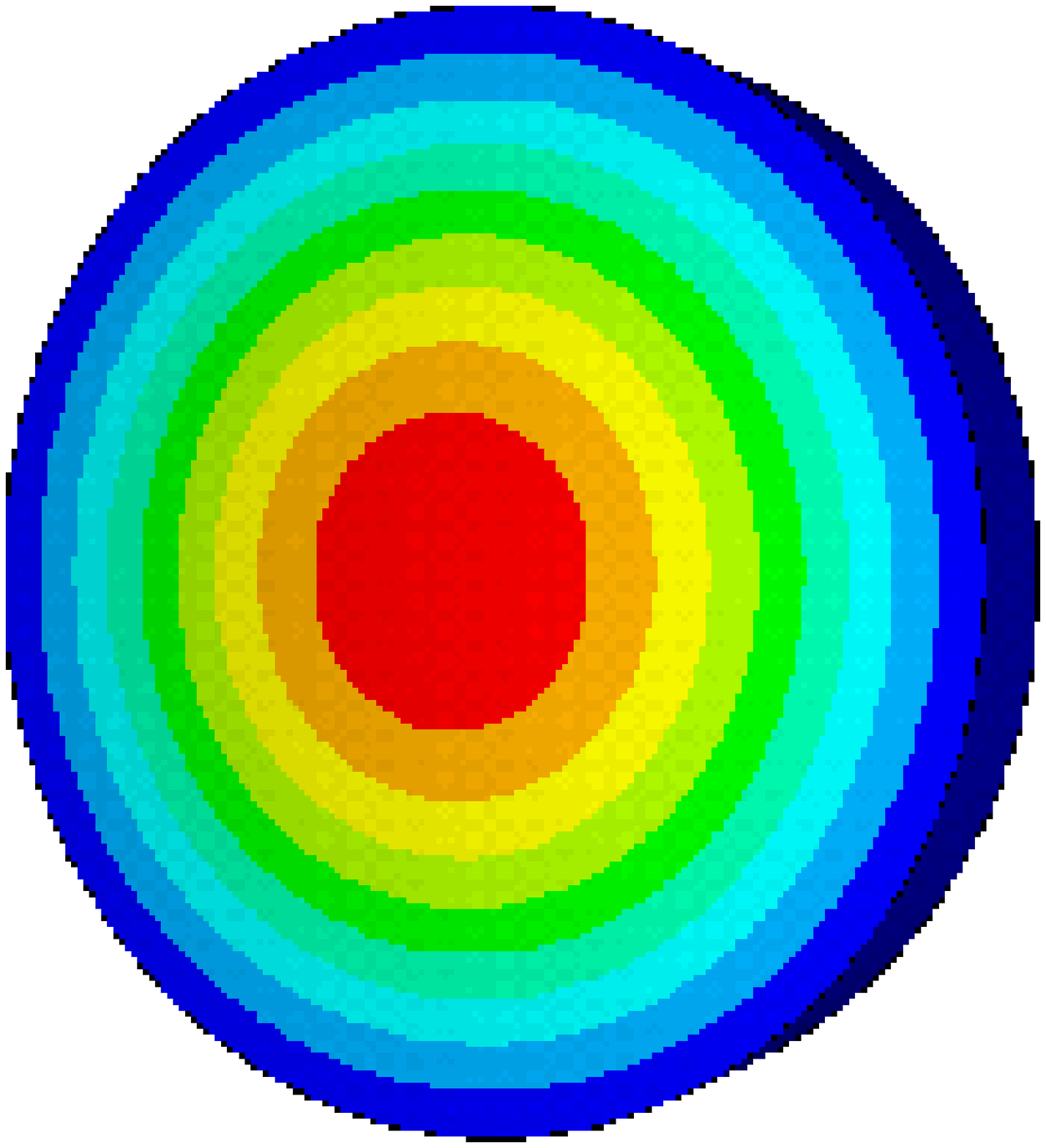}
c)~\includegraphics[width=.2\textwidth]{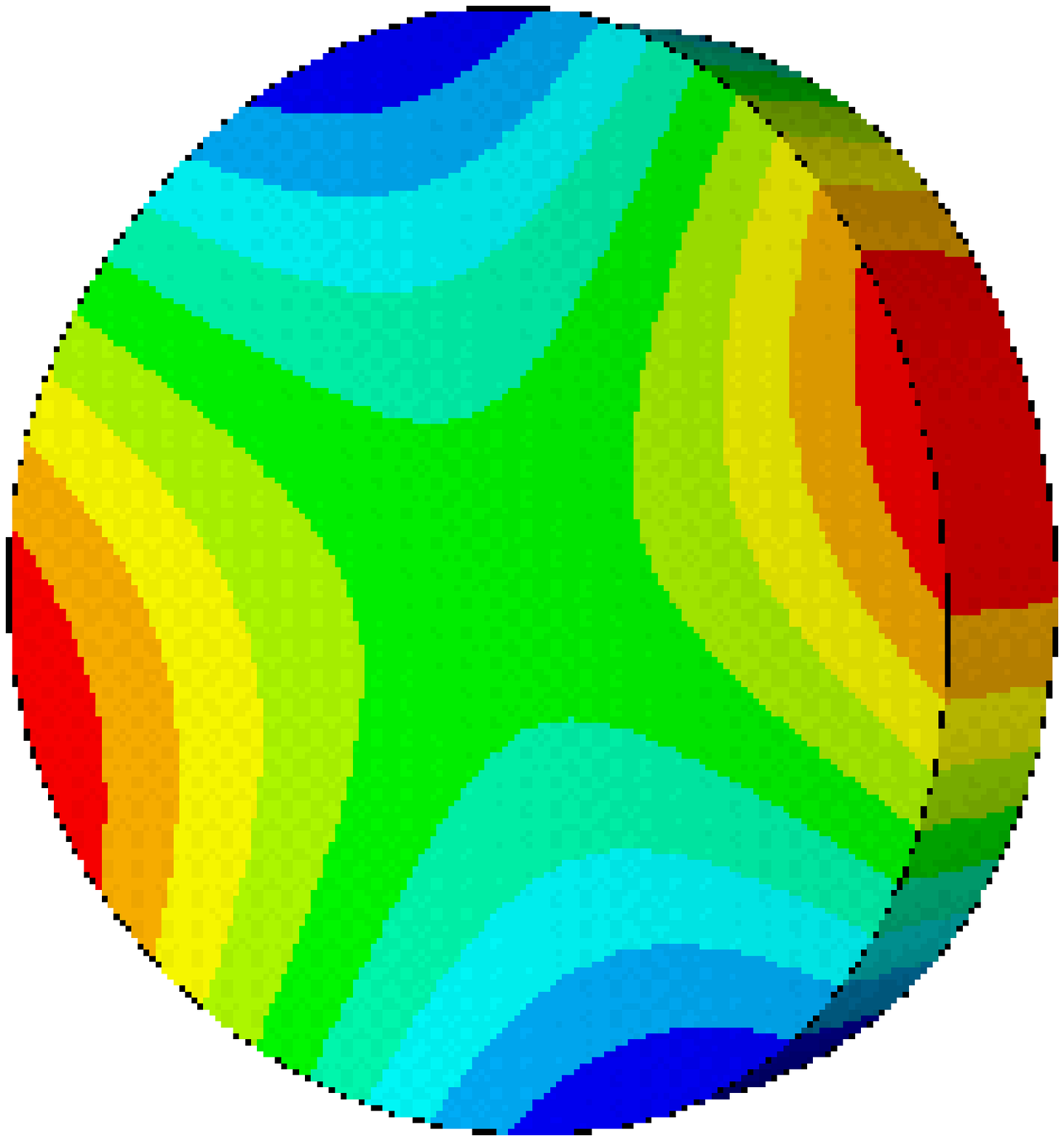} d)~\includegraphics[width=.2\textwidth]{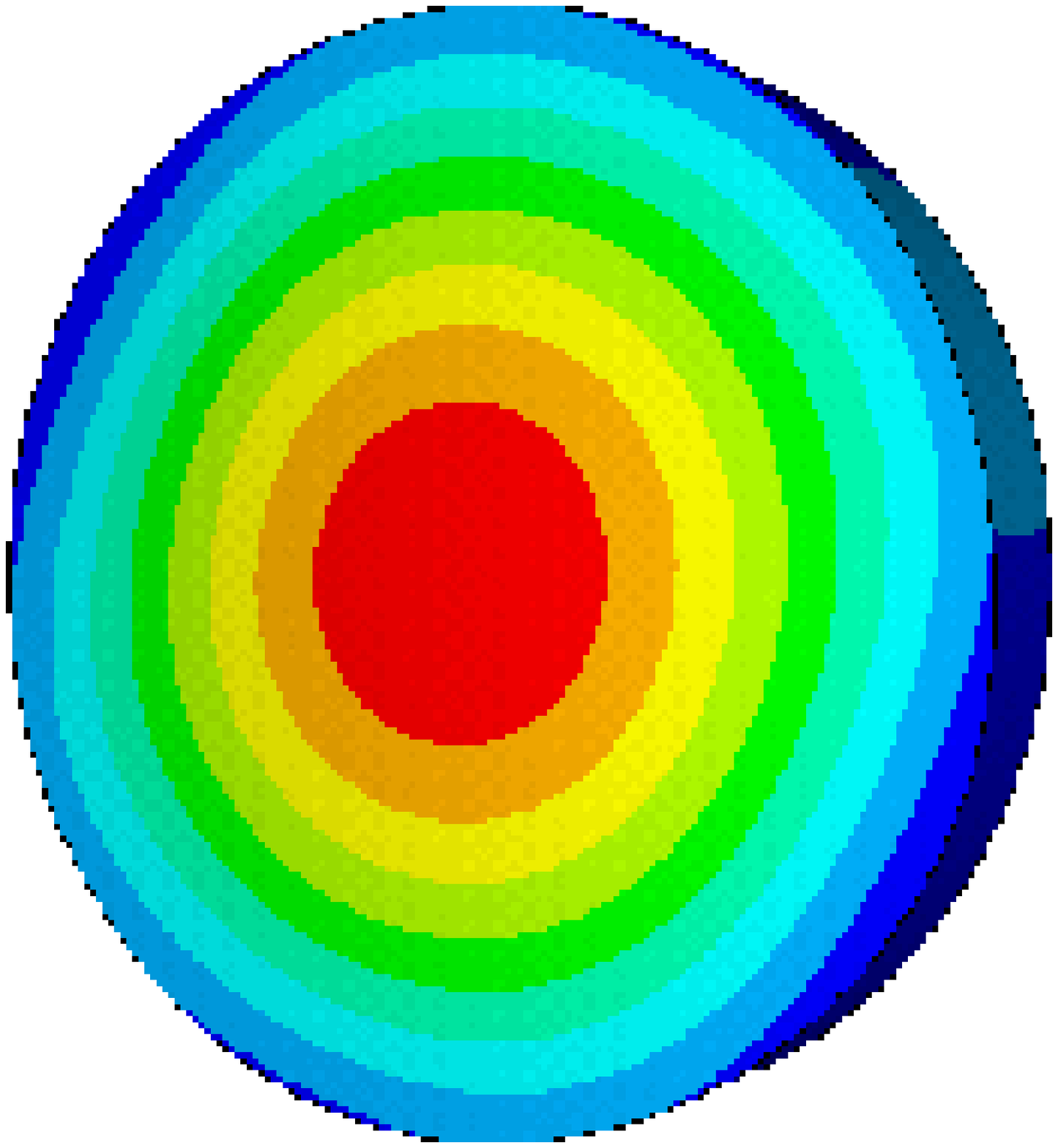}
\caption{Calculated mode shapes and frequencies of the  fused silica substrate a) 11548 Hz and b) 17101 Hz at
300\,K and the crystalline quartz substrate c) 10792 Hz and d) 16964 Hz at 300\,K. The cylinder axis is
aligned with the c-axis of crystal. Please note, that the asymmetric mode shape reflects the anisotropy of
the quartz crystal. For the numerical simulations the FEA based software ANSYS \cite{ans} was used. The
essential material parameters have been taken from SCHOTT Lithotec \cite{Schfs} for fused silica and from
Smakula et al. \cite{Sma1955} (density) and Bechmann \cite{Bec1958} (elastic coefficients) for crystalline
quartz.} \label{fig:modeshapes}
\end{figure}

In Fig.\,\ref{fig:fusedsilica} the reciprocals of the measured Q
factors are plotted versus temperature. The data  have been taken
on two substrates made of fused silica \cite{Schfs} and
crystalline quartz \cite{Schq}, both identical in shape (a cylinder 76.2\,mm in diameter and 12\,mm thick). The excited
mode shapes, visualized in Fig.\,\ref{fig:modeshapes}, were chosen
to be similar for both substrates. The mode shapes are not identical because the 
fused silica is isotropic, but the crystalline quartz is anisotropic and has different physical properties in the different directions.

For fused silica a large, broad damping peak occurs. No strong
dependency of the peak shape or height on the mode shape is
visible. The mechanical losses increase about 4 orders of
magnitude from room temperature to 30\,K finally resulting in
a value of $10^{-3}$ at 30\,K.\\
Fused silica is a material with high structural variation. The
structural unit, a silicon atom surrounded by 4 oxygen atoms in
the form of a tetrahedron, equals that of crystalline quartz. But
the variation in bond lengths and bond angles generates its
amorphous character. This variation also seems to be the cause for
the observed large, broad, damping peak
\cite{Hun1974,Hun1976,Hun1981,Wue1997}. In general one can assume
that a vast number of different bonding configurations with
different bonding energies exist.
Considering the simplest case, the energy landscape of two configurations can be described by two valleys of
an asymmetric double-well potential. The wall height between them is given by the stiffness of the directed
bonding forces and can be overcome by thermal energy.\\
Three possibilities of the microscopic origin of the mechanical losses have been proposed:\\
(1) transverse movement of oxygen atoms between two potential minima resulting in flipping of the bond \cite{And1955}\\
(2) motion of oxygen atoms in the direction of the bond between two potential minima \cite{Str1961}\\
(3) rotation of the $SiO_{4}$ tetrahedrons by small angles \cite{Vuk1972}.\\
Due to the amorphous character of fused silica a wide variation in
the height of the energy barrier occurs and therewith a
distribution in activation energies. An activation energy $E_{a}$
is needed to overcome the barrier between the two configurations.
In combination with the relaxation constant $\tau_{0}$ it
determines the temperature dependent relaxation time according to
an Arrhenius-like relation,
\begin{equation}
    \tau \: = \: \tau_{0} \cdot e^{\frac{E_{a}}{k_{B} \: T}}, \label{arrh}
\end{equation}
where $k_{B}$ is the Boltzmann constant and T the temperature. A
distribution in relaxation times requires integrating over all
contributions at a fixed temperature \cite{Now1961}. The
contribution of a single anelastic process is given by
Eq.\,\ref{zener}. For small losses ($\phi, \Delta \ll 1$) the
tangent is nearly linear,
\begin{equation}
    \phi \: \approx \: \Delta \: \frac{2\pi \cdot f \cdot \tau}{1+ \left(2\pi \cdot f \cdot \tau \right)^{2}}\,.
\end{equation}
The mechanical loss due to a superposition of loss processes with varying relaxation time can then be written
as
\begin{equation}
    \phi \: \approx \: \Delta \: \int^{\infty}_{-\infty} \Psi\left(\ln\tau\right) \frac{2\pi \cdot f \cdot \tau}
    {1+ \left(2\pi \cdot f \cdot \tau \right)^{2}} \: \textit{d} \: \ln\tau\,,
    \label{phi}
\end{equation}
with the following normalized distribution function
\begin{equation}
    \int^{\infty}_{-\infty} \Psi\left(\ln\tau\right) \: \textit{d} \: \ln\tau  \: = \: 1\,.
\end{equation}
The distribution is linearly related to $\ln\tau$, as $\ln\tau$ is linearly related to the activation energy
\cite{Now1961}. Assuming a Gaussian distribution of the barrier heights and therewith also of the activation
energies, the relaxation times are lognormal distributed,
\begin{equation}
    \Psi\left(\ln\left(\tau/\tau_{m}\right)\right) \: = \: \frac{1}{\beta \sqrt{\pi}}
    \exp\left[-\left(\ln\left(\tau/\tau_{m}\right)/\beta\right)^{2}\right] , \label{distri}
\end{equation}
where $\tau_{m}$ is the most likely value of $\tau$ and $\beta$ is
the half-width of the Gaussian distribution at the point where
$\Psi$ falls to $1/e$ of its maximum value. Since solely a
variation in activation energies and not in the relaxation
constant is assumed, $\beta$ is chosen to be temperature
dependent, $\beta = \beta_{c}/T$ \cite{Now1961}, where $\beta_{c}$
is a constant. The measured data recorded for the resonant
frequency of 17193\,Hz at 300\,K in Fig.\,\ref{fig:fusedsilica}
have been fit using Eq. \ref{phi} and \ref{distri} with the
following parameters: $\tau_{0} = 1\times10^{-13}s$, $\beta_{c} =
700\:1/K$ and $\tau_{m} = \tau_{0}\times
\exp\left(35.3\,meV/k_{B}T\right)$. The parameters used for
$\tau_{0}$ and $\tau_{m}$ are those reported in the work of
Hunklinger \cite{Hun1974}, and also successfull describe our
measurement data between 50\,K and 100\,K, see the dashed line in
Fig.\,\ref{fig:fusedsilica}.
Differences to the measured values at other temperatures are possibly due to the following mechanisms. For
temperatures above 225\,K, phonon-phonon interactions are plausible. Phonon-phonon interactions are expected
to play a similarly important role in both, fused silica and crystalline quartz, which is indeed found in
(Fig.\,\ref{fig:fusedsilica}). The deviations between 100\,K and
225\,K and below 50\,K are assumed to be due to other loss mechanisms. At 50\,K as well as at about 100\,K, a small kink in the curve is visible indicating additional damping peaks. Below 50\,K a damping peak
has been observed by Hunklinger \cite{Hun1974} at far higher resonant frequencies (507 and 930\,MHz). Due to
the much higher frequencies the damping peaks were clearly separated from each other and observed as distinct
peaks. Note that a loss peak caused by relaxation processes moves with decreasing frequency to lower temperatures. Peaks having different parameters entering the relation for the relaxation time are not equally
shifted. Thus, the peak observed by Hunklinger is likely to appear in our measurements below 50~K partially covered by the damping peak due to structural variations since the resonant frequencies are lower. According to the work of Hunklinger \cite{Hun1974} it is caused by relaxation
processes of the two-level-systems by resonant absorption or emission of thermal phonons.
The origin of the damping peak at about 150\,K is not known to our knowledge. An analysis of this problem
would require a clearer separation of this damping peak from our fitted damping peak at about 80\,K. This
could be achieved by measurements at lower frequencies which are, however, experimentally rather challenging.

For crystalline quartz the temperature spectrum of the measured Q
factors looks rather different. Below room temperature the damping
slightly decreases with decreasing temperature. However, several
damping peaks are visible with heights depending on the resonant
frequency. In particular the temperature region from 30\,K to
50\,K is dominated by a rather narrow
distinct damping peak. Below 15\,K the losses strongly decrease. We found Q values of up to $7\times10^7$.\\
The origin of the losses in crystalline quartz can be found in interactions of the excited acoustic waves
with thermal phonons of the crystal and with alkali ions which were introduced during the growth
\cite{Fra1964}. Relaxations of the perturbed states to new equilibria lead to removal of oscillation energy.
However, the defect induced damping peaks can be reduced or removed by sweeping (electrodiffusion) or by
irradiating the crystal \cite{Mar1984}. For a review on loss mechanisms in crystalline quartz see the work of
Zimmer et al. \cite{Zim2007}.
In contrast to fused silica, crystalline quartz shows a much
narrower damping peak because almost no variations in bond lengths
and bond angles occur. Quite generally most relaxation processes
in crystalline materials depend on parameters with low variations
and result in narrow damping peaks. Crystalline quartz is however
piezoelectric thus the possibility of excess damping associated
with this piezoelectric behaviour is an area which require
furthers investigation. The crystalline structure also causes an
anisotropy of the material which explains why the damping depends
on the direction of the motion, i.e. on the mode shape.

The advantages of crystalline quartz over its amorphous
counterpart as discussed here are evident in other crystalline
materials. Crystalline silicon shows very high Q factors at
cryogenic temperatures \cite{MLGHDG78} in addition to having other
highly desirable thermo-mechanical properties \cite{Row2003} and
may be used as a test mass material if future IGWDs operate at a
laser wavelengths of 1550\,nm, or employ non-transmissive
topologies \cite{Bye1990,Dre1995,Sun1998,Bun2004,Bun2006,Sch2006,Ronny07}. Another
crystalline high Q material is sapphire \cite{Bra1986,sapphire99},
which is already used in the Japanese cryogenic IGWD prototype
\cite{Miy2006}.

\section{Conclusions}

We confirmed an increasing mechanical loss of fused silica from
room temperature down to a maximum at 30\,K measured at resonant
frequencies of slightly above 10\,kHz. The emergence of broad loss
peaks is a general behaviour of amorphous solids and cannot be
circumvented by higher purities. Therefore, fused silica is not
suitable as a material for the test masses of cryogenic
gravitational wave detectors. Crystalline quartz, however has more
promise. Q factors of up to 7$\times$10$^7$ were measured on our
quartz test mass, suggesting crystalline quartz may merit further
study for IGDWs operated at the laser wavelength of 1064\,nm. We
note that for frequencies in the gravitational wave detection band
around 200\,Hz the measured crystalline quartz curves in
Fig.\,\ref{fig:fusedsilica} need to be shifted to the left by 10
to 20\,K. This is a result of the frequency dependence of the loss
factor according to Eq.\,(\ref{zener}). Assuming frequency
independent relaxation strengths and times for four independent
damping peaks, as found below 150\,K in the crystalline quartz
investigated here, 50\,K might turn out to be a preferred
temperature for a low thermal noise performance in the
gravitational wave detection band. We note that some of the
measured losses which arise due to defect relaxations can possibly
be reduced by higher purity, sweeping or irradiating the crystal.

\ack  This work was supported by the DFG (Deutsche
Forschungsgemeinschaft) under contract SFB Transregio 7. We also
wish to thank the University of Glasgow, STFC and the Leverhulme
Trust for support and our colleagues in the GEO project and LIGO
Scientific Collaboration for their interest in this work.

\end{document}